\documentclass[prd,twocolumn,superscriptaddress,showpacs,nofootinbib,preprintnumbers]{revtex4}

\usepackage{amsmath}
\usepackage{amsfonts}
\usepackage{graphicx}
\usepackage{dcolumn}

\def\be{\begin{equation}}
\def\ee{\end{equation}}
\def\ba{\begin{eqnarray}}
\def\ea{\end{eqnarray}}
\def\bs{\begin{subequations}}
\def\es{\end{subequations}}

\newcommand{\rd}{{\rm d}}

\newcommand{\vp}{\varphi}

\usepackage{color}

\begin{document}

\title{General analytic formulae for attractor solutions of 
scalar-field dark energy models \\
and their multi-field generalizations}

\author{Shinji Tsujikawa}
\email{shinji@nat.gunma-ct.ac.jp}
\affiliation{Department of Physics, Gunma National College of
Technology, Gunma 371-8530, Japan}
\date{\today}

\begin{abstract}
    
We study general properties of attractors for scalar-field
dark energy scenarios which possess cosmological scaling solutions.
In all such models there exists a scalar-field dominant solution with an energy 
fraction $\Omega_{\phi}=1$ together with a scaling solution.
A general analytic formula is given to derive fixed points
relevant to dark energy coupled to dark matter. 
We investigate the stability of fixed 
points without specifying the models of dark energy in the presence of 
non-relativistic dark matter and provide a general 
proof that a non-phantom scalar-field 
dominant solution is unstable when a stable scaling solution 
exists in the region $\Omega_{\phi}<1$.
A phantom scalar-field dominant fixed point is found to be 
classically stable. We also generalize the analysis to the case of 
multiple scalar fields and show that for a non-phantom scalar field
assisted acceleration always occurs for all 
scalar-field models which have scaling solutions.
For a phantom field the equation of state approaches that of 
cosmological constant as we add more scalar fields.

\end{abstract}


\maketitle

\section{Introduction}

Dark energy is one of the most serious mysteries 
in modern cosmology. The observations of 
supernova (SN) Ia \cite{SNIa} 
have continuously confirmed that about 70\% of the total energy 
density of universe consists of unknown energy which leads 
to an accelerated expansion of the universe.
The existence of dark energy has been also supported from 
the observations of cosmic microwave background (CMB)
\cite{CMB}, large-scale structure (LSS) \cite{LSS} and baryon oscillation 
experiments \cite{baryon} (see also Refs.~\cite{obser}).

While cosmological constant can be a natural candidate of dark 
energy, this suffers from a severe fine-tuning problem if it 
originates from particle physics 
(see Refs.~\cite{rev1,rev2,rev3,rev4} for review).
Instead of sticking to cosmological constant, a host of alternative
scenarios have been proposed-- ranging from scalar-field models,
Chaplygin gas \cite{Chap}, braneworld \cite{brane} and 
modified gravity scenarios in General Relativity \cite{Capo}. 
Since scalar fields are natural 
ingredients in particle physics, it is an interesting challenge 
for theoreticians to construct viable dark energy models using them.
A number of scalar-field  models have been 
studied as candidates of dark energy-- an incomplete 
list includes quintessence \cite{quin}, 
k-essence \cite{kes}, tachyons \cite{tachyon}, 
phantoms \cite{phantom} and (dilatonic) ghost 
condensates \cite{dilaton,PT}.
These can at least alleviate the fine-tuning problem of 
cosmological constant because of their dynamical nature.

A general guidance to obtain viable dark energy models
is that the energy density of it is ``hidden'' during radiation 
and matter dominant eras and comes out near to the 
present epoch. If one uses the property of cosmological 
scaling solutions \cite{Fer,CLW,scaling,Maco}, 
the energy density of scalar fields 
decreases proportionally to that of the background fluid
(radiation or matter) independently of the initial 
conditions \cite{Nelson}.
The system can exit from the scaling regime to give rise 
to the present accelerated expansion if the potential of the 
field becomes shallow \cite{Nelson,SW} or if the coupling $Q$ 
between dark energy and dark matter becomes 
important \cite{Luca1}.
Especially in the latter case there is an intriguing possibility 
that the present universe is a scaling attractor with a constant 
dark energy fraction $\Omega_\phi \simeq 0.7$ \cite{Luca2}.
Thus the presence of scaling solutions plays an important 
role for the model-building of dark energy.

It is well known that exponential potentials give rise to scaling 
solutions for an ordinary scalar field \cite{CLW}. 
The form of the potentials
which generate scaling solutions depends on the models of 
dark energy. For example in the case of a tachyon field the 
corresponding potential is given by 
$V(\phi)=V_0 \phi^{-2}$ \cite{Abramo,Laz,CGST}. 
A general algorithm to obtain potentials corresponding to 
scaling solutions is given in Refs.~\cite{PT,TS}.
Interestingly the scalar-field Lagrangian density for the existence of 
scaling solutions is restricted to be in a compact form 
$p=Xg(Xe^{\lambda \phi})$ \cite{PT}, 
where $X$ is the kinematic term of 
a scalar field $\phi$ and $g$ is an arbitrary function.
This analysis can be extended to the more general case in which 
the Hubble parameter $H$ has a dependence $H^2 \propto \rho^n$
with $\rho$ being the total energy density \cite{TS}.
By using the formula in Ref.~\cite{TS} we can easily obtain 
potentials which generate scaling solutions once we specify 
dark energy models.

In addition to scaling solutions there exists another important 
fixed point which corresponds to the scalar-field dominant 
solution characterized by $\Omega_\phi=1$.
In fact, in the case of an ordinary scalar field with 
an exponential potential, this solution was used in 
Ref.~\cite{Nelson} to obtain a late-time attractor 
with an accelerated expansion 
after the system exits from 
a scaling regime during radiation and matter dominant eras. 
In Ref.~\cite{BNST} it was shown that the scalar-field dominant 
solution also exists for tachyons and dilatonic ghost condensates.
In this paper we will show that this fixed point always exists
for all scalar-field models which possess scaling solutions.
We shall also provide a general analytic formula to derive
critical points for both the scalar-field dominant solution and 
the scaling solution.

The stability of critical points against perturbations was analyzed 
in Ref.~\cite{BNST} by specifying three models of dark energy.
We shall show that the stability analysis can be analytically done
for all dark energy models which have scaling solutions.
We will generally prove that the non-phantom scalar-field 
dominant solution is unstable when the stable scaling solution 
exists in the region $\Omega_{\phi}<1$.
It will be also shown that the phantom scalar-field dominant
solution is classically stable.
This analysis is useful when we construct viable 
dark energy models.

There is another interesting possibility to give rise to an 
inflationary solution by using multiple scalar fields.
In the case of an ordinary scalar field with an exponential potential
many fields can cooperate to sustain inflation even 
if none is able to do individually \cite{LMS}. 
This assisted inflation scenario was applied to dark energy 
in Refs.~\cite{Coley,KLT}.
The field energy density mimics that of the background fluid
during radiation and matter dominant eras because of the scaling 
property, which is followed by the scalar-field dominant solution 
as more fields join to sustain an accelerated expansion.
In this paper we shall show that for a non-phantom field 
this assisted behavior always
happens for {\it all} scalar-field dark energy models which 
have scaling solutions.

\section{Conditions for the existence of scaling solutions}

In this section we shall review the Lagrangian for the existence
of scaling solutions presented in Refs.~\cite{PT,TS}.
Let us start with the following 4-dimensional action
\begin{eqnarray}
\label{action}
S=\int {\rm d}^4 x \sqrt{-g} \left[\frac{R}{2}
+ p(X, \phi)\right]+S_m (\phi)\,,
\end{eqnarray}
where $R$ is a scalar curvature and
$X =-g^{\mu\nu} \partial_\mu \phi \partial_\nu \phi /2$ 
is a kinematic term of a scalar field $\phi$. 
$S_m$ is an action for a barotropic fluid
which is generally dependent on $\phi$.
The reduced Planck mass $M_{\rm pl}$
is set to be unity.
The above action covers most of scalar-field
dark energy models.

The pressure density and  the energy density of the field 
are given, respectively, by 
\begin{eqnarray}
\label{prho}    
p_{\phi}=p\,,~~~
\rho_{\phi}=2X \frac{\partial p}{\partial X}-p' \,.
\end{eqnarray}
We introduce a scalar charge $\sigma$ defined by 
\begin{eqnarray}
\sigma=-\frac{1}{\sqrt{-g}}
\frac{\delta S_m(\phi)}{\delta \phi}\,,
\end{eqnarray}
which corresponds to the coupling between 
the field $\phi$ and the barotropic fluid.
This coupling can lead to an accelerated expansion
even when scaling solutions can not do so 
in the absence of the interaction \cite{Luca1}.
See Refs.~\cite{couple} for recent works 
about the interacting dark energy.
A similar coupling arises in scalar-tensor 
theories \cite{Luca1,stensor} and 
in neutrino models of dark energy 
in which the mass of neutrinos depends on a
scalar field $\phi$ \cite{neutrino}.

In a flat Friedmann-Robertson-Walker (FRW) background 
with a scale factor $a$, 
the continuity equation for the field $\phi$ is given by 
\begin{eqnarray}
\label{geneeq1}
\dot{\rho}_{\phi}+3H(1+w_\phi)\rho_{\phi}
=-Q\rho_m \dot{\phi}\,,
\end{eqnarray}
where a dot represents a derivative with respect to 
cosmic time $t$, 
$Q \equiv \sigma/\rho_{m}$, 
$w_{\phi} \equiv p_{\phi}/\rho_{\phi}$ and 
$H \equiv \dot{a}/a$ is the Hubble rate.
Meanwhile the fluid energy density, $\rho_{m}$, satisfies
\begin{eqnarray}
\label{geneeq2}
\dot{\rho}_{m}+ 3H(1+w_m) \rho_m =  
Q \rho_m \dot{\phi}\,.
\end{eqnarray}
In what follows the equation of state of 
a barotropic fluid, $w_{m} \equiv p_{m}/\rho_m$, 
is assumed to be constant.

The Hubble parameter $H$ obeys the Friedmann equation
\begin{eqnarray}
\label{Hubble}
3H^2=\rho_{\phi}+\rho_{m}\,.
\end{eqnarray}
The fractional densities of $\rho_{\phi}$ and $\rho_m$ are 
defined by 
\begin{eqnarray}
\label{Omepm}    
\Omega_\phi \equiv \frac{\rho_{\phi}}{3H^2}\,,~~~
\Omega_m \equiv \frac{\rho_m}{3H^2}\,.
\end{eqnarray}
These satisfy the relation $\Omega_\phi+ \Omega_m =1$ 
from Eq.~(\ref{Hubble}).
We note that reconstruction equations suitable 
for the comparison to SN Ia observations
were obtained in Ref.~\cite{shinji05} for the action 
(\ref{action}).

Cosmological scaling solutions are characterized 
by the condition 
\begin{eqnarray}
\label{scaling}
\rho_{\phi}/\rho_{m}={\rm const}\,\,\,(\neq 0)\,.
\end{eqnarray}
In this case $\Omega_\phi$ and $\Omega_{m}$ are 
constant from Eq.~(\ref{Omepm}). 
We also assume that $Q$ and $w_{\phi}$ 
do not vary in the scaling regime.
Note that the constant coupling $Q$ appears in theories
in which the mass of dark matter or neutrinos 
depends exponentially on the scalar field, 
see e.g., Refs.~\cite{Comelli,neutrino}.

{}From the condition (\ref{scaling})
we find $({\rm log} \rho_\phi)^{\bullet}
=({\rm log} \rho_m)^{\bullet}$.
Then we obtain the following relation from
Eqs.~(\ref{geneeq1}) and (\ref{geneeq2}):
\begin{equation}
\label{dphi}
\frac{\rd \phi}{\rd N} = 
\frac{3\Omega_\phi}{Q}(w_m - w_\phi)
={\rm const}\,,
\end{equation}
where $N \equiv \int H {\rm d}t$ is the number of 
$e$-foldings. This gives the following scaling behavior:
\begin{equation}
\label{sca}
\frac{\rd {\rm ln} \rho_{\phi}}{\rd N}=
\frac{\rd {\rm ln} \rho_m}{\rd N}=
-3(1+w_{\rm eff})\,,
\end{equation}
where the effective equation of state is
\begin{equation}
\label{ws}
w_{\rm eff} \equiv
\frac{w_{\phi}\rho_{\phi}+w_{m}\rho_{m}}
{\rho_{\phi}+\rho_{m}}
=w_m+\Omega_\phi (w_\phi-w_m)\,.
\end{equation}
This expression of $w_{\rm eff}$ can be used 
irrespective of the fact that scaling solutions exist or not.

The definition of $X$ gives the relation
\begin{equation}
\label{Xdef}
2 X = H^2 \left(\frac{\rd \phi}{\rd N}\right)^2 
\propto \, H^2 \propto 
(\rho_{\phi}+\rho_{m})\,,
\end{equation}
where we used Eq.~(\ref{dphi}).
Hence the scaling property of $X$ is the same 
as $\rho_{\phi}$ and $\rho_m$, which leads
to the relation
\begin{equation}
\label{X2eq}
\frac{\rd {\rm ln} X}{\rd N}=-3(1+w_{\rm eff})\,.
\end{equation}
The pressure density $p_{\phi}=w_\phi \rho_{\phi}$ 
scales in the same way as $\rho_{\phi}$, i.e.,
$\rd {\rm ln}\,p_{\phi}/\rd N=-3(1+w_{\rm eff})$.
Using this relation together with Eqs.~(\ref{dphi}) 
and (\ref{X2eq}) we find
\begin{equation}
\label{pform}
\frac{\partial \ln p_{\phi}}{\partial \ln X} -
\frac{1}{\lambda} 
\frac{\partial \ln p_{\phi}}{\partial \phi} = 1\,,
\end{equation}
where
\begin{equation}
\label{lam2}
\lambda\,  \equiv\,
Q \frac{1+w_m - \Omega_\phi (w_m - w_\phi)}
{\Omega_\phi (w_m-w_\phi)}\,.
\end{equation}

{}From Eq.~(\ref{pform}) the existence of scaling solutions
restricts the Lagrangian density in  the form:
\begin{equation}
\label{scap}
p(X, \phi) = X\,
g\left(X e^{\lambda \phi}\right)\,,
\end{equation}
where $g$ is an arbitrary function in terms of 
$ Y \equiv X e^{\lambda \phi}$. 
This expression was first derived in Ref.~\cite{PT}
and was extended to the more general case in which 
the Friedmann equation is given by 
$H^2 \propto (\rho_\phi+\rho_m)^n$ \cite{TS}.
The quantity $Y$ is conserved along the scaling 
solution, i.e., 
$Y=X e^{\lambda \phi} = {\rm constant}$.

Although we derived the expression (\ref{scap}) under 
the assumption that $Q$ is a constant, this can be generalized
to the case in which $Q$ depends on the field $\phi$.
This case will be presented in elsewhere \cite{ASappear}.
We also note that the r.h.s. of Eq.~(\ref{dphi}) 
is kept to be a constant for $Q \to 0$, 
since $w_{\phi} \to w_{m}$ for scaling solutions in this limit. 
Thus the above discussion for the derivation of 
Eq.~(\ref{scap}) is valid even for $Q=0$
by taking the limit $Q \to 0$.

Equations (\ref{dphi}), (\ref{Xdef}) and (\ref{lam2})
give the relation
\begin{equation}
\label{scarelation}    
3H^2=\frac{2(Q+\lambda)^2}{3(1+w_m)^2}X\,.
\end{equation}
{}From Eqs.~(\ref{ws}) and (\ref{lam2}) we find
\begin{equation}
\label{weff2}    
w_{\rm eff}=\frac{w_{m}\lambda-Q}{Q+\lambda}\,.
\end{equation}
When $Q=0$ this reduces to $w_{\rm eff}=w_{m}$. 
Hence an accelerated expansion is 
not possible for a non-relativistic fluid ($w_{m}=0$).
However the presence of the coupling $Q$ leads to 
an accelerated expansion ($w_{\rm eff}<-1/3$)
for $Q>(3w_{m}+1)\lambda/2$.

If we choose the function $g(Y)=1-c/Y$ in Eq.~(\ref{scap}), 
we obtain the Lagrangian density for an ordinary 
scalar field with an exponential potential: 
$p=X-ce^{-\lambda \phi}$ \cite{Fer,CLW}.
The function $g(Y)=-1+cY$ gives the dilatonic ghost 
condensate model with Lagrangian density 
$p=-X+cX^2$ \cite{PT}.
When we choose the function 
$g(Y)=-c\sqrt{1-2Y}/Y$ together with the introduction of 
a new field $\vp=(2/\lambda)e^{\lambda \phi/2}$, we obtain 
the tachyon Lagrangian $p=-V(\vp) \sqrt{1-\dot{\vp}^2}$
with an inverse square potential 
$V(\vp)=4c/(\lambda^2 \vp^2)$ \cite{Laz,CGST}.

\section{Fixed points}

In the previous section we showed that the Lagrangian (\ref{scap})
possesses scaling solutions. 
There exist other fixed points for the system (\ref{scap}). 
We shall derive fixed points relevant to dark energy and 
analyze the stability of them against perturbations.
Note that critical points were derived for three dark energy models
in Ref.\,\cite{BNST} by specifying the functional form of $g(Y)$.
In this paper we show that this is possible for all scalar-field 
models that possess scaling solutions.
In what follows we shall consider the case of 
positive values of $Q$ and $\lambda$.

{}From Eqs.~(\ref{prho}) and (\ref{geneeq1}) together with 
the Lagrangian density (\ref{scap}),
we find that the scalar field obeys the equation of motion:
\begin{equation}
\label{ddphi}
\ddot{\phi}+3HAp_{X}\dot{\phi}+
\lambda X \left[1-A(g+2Yg') \right]
=-AQ\rho_m\,.
\end{equation}
where a prime represents a derivative 
with respect to $Y$ and
\begin{eqnarray}
& &A(Y) \equiv \left[g(Y)+5Yg'(Y)+2Y^2g''(Y)\right]^{-1}\,, \\
& &p_{X}(Y) \equiv \frac{{\rm d}p}{{\rm d}X}
=g(Y)+Yg'(Y)\,.
\end{eqnarray}
Combining Eqs.~(\ref{geneeq1}) and (\ref{geneeq2}) with 
(\ref{Hubble}), we find
\begin{eqnarray}
\label{dotH}
\dot{H}=-Xp_{X}-\frac12 (1+w_m) \rho_{m}\,.
\end{eqnarray}

We introduce the following dimensionless quantities which 
are useful to study the dynamical system:
\begin{eqnarray}
\label{xydef}    
x \equiv \frac{\dot{\phi}}{\sqrt{6}H}\,,~~~
y \equiv \frac{e^{-\lambda \phi/2}}{\sqrt{3}H}\,.
\end{eqnarray}
The variables $x$ and $y$ can be regarded as a ``kinematic'' term 
and a ``potential'' term, respectively. 
Since we do not consider a contracting universe, 
$y$ is positive from its definition.
The variable $Y=Xe^{\lambda \phi}$ is expressed
in terms of $x$ and $y$:
\begin{eqnarray}
\label{Yre}    
Y=x^2/y^2\,.
\end{eqnarray}

Equation (\ref{Hubble}) gives the constraint equation
$\Omega_\phi+\Omega_m=1$ with 
\begin{eqnarray}
\Omega_{\phi} \equiv x^2(g+2Yg')\,.
\end{eqnarray}
The equation of state for the scalar field 
$\phi$ is given by 
\begin{equation}
w_{\phi}=\frac{g}{g+2Yg'}\,.
\end{equation}
It may be useful to notice the relation 
\begin{equation}
\label{Omew}    
\Omega_{\phi} w_{\phi}=gx^2\,,\quad
w_{\phi}=-1+\frac{2x^2}{\Omega_{\phi}}p_{X}\,.
\end{equation}
This shows that the field behaves as a phantom 
($w_{\phi}<-1$) for $p_{X}<0$
provided that $\Omega_{\phi}>0$.

{}From Eqs.~(\ref{ddphi}) and (\ref{dotH}) 
we get the following autonomous equations for
$x$ and $y$:
\begin{eqnarray}
\label{xeq}
\hspace*{-1.3em}
\frac{{\rm d}x}{{\rm d}N} &=&
\frac{3x}{2} \left[1+gx^2-w_{m} (\Omega_{\phi}
-1)-\frac{\sqrt{6}}{3} \lambda x \right] \nonumber \\
& &+\frac{\sqrt{6}A}{2} \left[ (Q+\lambda)
\Omega_{\phi}-Q-\sqrt{6}(g+Yg')x \right], \\
\label{yeq}
\hspace*{-1.3em}
\frac{{\rm d}y}{{\rm d}N} &=&
\frac{3y}{2} \left[1+gx^2-w_{m} (\Omega_{\phi}
-1)-\frac{\sqrt{6}}{3} \lambda x \right]\,.
\end{eqnarray}
We can obtain fixed points of the system 
by setting ${\rm d}x/{\rm d}N=0$ and 
${\rm d}y/{\rm d}N=0$.
{}From Eq.~(\ref{yeq}) we find that 
$y=0$ corresponds to one of the fixed points.
However this is irrelevant to dark energy, 
since we need the contribution of a ``potential''
term to give rise to an accelerated expansion.
In addition this fixed point is unstable against 
perturbations \cite{BNST}.

Then the fixed points we are interested in 
satisfy the following equations:
\begin{eqnarray}
\label{basic1}    
& & \sqrt{6} \lambda x=
3\left[1+gx^2-w_{m} (\Omega_{\phi}-1)\right]\,, \\
\label{basic2}  
& & \sqrt{6}(g+Yg')x=(Q+\lambda)
\Omega_{\phi}-Q\,.
\end{eqnarray}
Using Eq.~(\ref{Omew}) we find that 
$g+Yg'=\Omega_{\phi}(1+w_{\phi})/2x^2$.
Then Eqs.~(\ref{basic1}) and (\ref{basic2})
can be written in the form:
\begin{eqnarray}
\label{xexpress}    
x&=& \frac{\sqrt{6}[1+(w_\phi- w_{m})\Omega_{\phi}
+w_{m}]}{2\lambda} \\
&=& \frac{\sqrt{6}(1+w_{\phi})\Omega_{\phi}}
{2[(Q+\lambda)\Omega_{\phi}-Q]}\,,
\end{eqnarray}
which leads to 
\begin{eqnarray}
(\Omega_{\phi}-1) \left[ (w_\phi- w_{m})
(Q+\lambda)\Omega_{\phi}+Q(1+w_{m})\right]=0\,.
\end{eqnarray}

Hence we find two cases:
\begin{itemize}
\item (i)
A scalar-field dominant solution with 
\begin{eqnarray}
\label{fixedi}    
\Omega_{\phi}=1\,.
\end{eqnarray}
\item (ii)
A scaling solution with 
\begin{eqnarray}
\label{scaome}    
\Omega_{\phi}=
\frac{(1+w_m)Q}{(w_{m}-w_{\phi})(Q+\lambda)}\,.
\end{eqnarray}
\end{itemize}

In the case of an ordinary scalar field, for example, 
the solution (i) corresponds to the 4-th fixed point in Table I 
of Ref.~\cite{CLW}, whereas the solution (ii) corresponds 
to the 5-th fixed point. 
The above discussion shows that similar fixed points exist
for all scalar-field dark models which possess scaling 
solutions. In what follows we shall study the properties of
the critical points in more details.

\subsection{Scalar-field dominant solutions}

When $\Omega_\phi=1$, Eqs.~(\ref{Omew}) 
and (\ref{basic1}) give the equation of state:
\begin{eqnarray}
\label{eqw}    
w_{\phi}=-1+\frac{\sqrt{6}\lambda}{3}x\,.
\end{eqnarray}
In this case the effective equation of state of the 
system is given by $w_{\rm eff}=w_{\phi}$ by 
Eq.~(\ref{ws}).
The condition for an accelerated expansion, 
$w_{\rm eff}<-1/3$, corresponds to 
\begin{eqnarray}
\label{acc}    
\lambda x< \frac{\sqrt{6}}{3}\,.
\end{eqnarray}

{}From Eq.~(\ref{basic2}) we find that $x$ is given by 
$x=\lambda/\sqrt{6}p_X$. Then $w_\phi$ is 
written in the form 
\begin{eqnarray}
\label{wsingle}    
w_{\phi}=-1+\frac{\lambda^2}{3p_X}\,.
\end{eqnarray}
Hence one has $w_\phi>-1$ for $p_X>0$ 
and $w_\phi<-1$ for $p_X<0$.
For an ordinary scalar field with an exponential potential, 
the Lagrangian density is given by $p=X-ce^{-\lambda \phi}$.
Since $p_{X}=1$ in this case the equation of 
state is $w_{\phi}=-1+\lambda^2/3$.
This agrees with the result obtained in Ref.~\cite{CLW}.
In the case of a phantom field with an exponential potential
($p=-X-ce^{-\lambda \phi}$), we obtain 
$w_{\phi}=-1-\lambda^2/3$.

{}From Eqs.~(\ref{basic1}) and (\ref{basic2}) we find
\begin{eqnarray}
\label{gre1}    
g(Y)=\frac{\sqrt{6}\lambda x-3}{3x^2}\,,~~~
Yg'(Y)=\frac{6-\sqrt{6}\lambda x}{6x^2}\,.
\end{eqnarray}
Once we specify a scalar-field dark energy model, i.e., 
the function $g(Y)$, we get $Y$ and $x$ from 
Eq.~(\ref{gre1}).
Using the relation (\ref{Yre}) we obtain the scalar-field
dominant fixed point $(x, y)$.
The equation of state is known by using Eq.~(\ref{eqw}).

Let us consider the ordinary (phantom) scalar field with $g(Y)$
given by $g(Y)=\epsilon- c/Y$ 
($\epsilon=+1$ for the ordinary field and 
$\epsilon=-1$ for the phantom).
{}From Eq.~(\ref{gre1}) we find 
$Y=c\lambda^2/(6-\epsilon \lambda^2)$ and 
\begin{eqnarray}
x=\frac{\lambda}{\sqrt{6}\epsilon}\,,~~~
y=\sqrt{\frac{1}{c} \left( 1-\epsilon 
\frac{\lambda^2}{6}\right)}\,.
\end{eqnarray}
This agrees with what was obtained in Refs.~\cite{CLW,BNST}.

In the case of dilatonic ghost condensate with
$g(Y)=-1+cY$, Eq.~(\ref{gre1}) leads to 
\begin{eqnarray}
\label{xcY}    
x_{1, 2}=-\frac{\sqrt{6}\lambda f_\pm (\lambda)}{4}\,,\quad
cY_{1, 2}=\frac12+\frac{\lambda^2 f_{\mp}(\lambda)}{16}\,,
\end{eqnarray}
where 
\begin{eqnarray}
f_{\pm} (\lambda)=1 \pm \sqrt{1+16/(3\lambda^2)}\,.
\end{eqnarray}
This result agrees with what was obtained in Ref.~\cite{PT,BNST}.
We have two fixed points in this model.
Since $f_{+}(\lambda)>0$ and $f_{-}(\lambda)<0$, 
the critical point $(x_1, y_1)$ corresponds to a phantom 
one with $w_\phi<-1$ from Eq.~(\ref{eqw}), 
whereas the point ($x_{2}, y_{2}$) satisfies $w_\phi>-1$.
In the latter case the condition (\ref{acc}) for an 
accelerated expansion corresponds to $\lambda<\sqrt{6}/3$.
The phantom divide is characterized by the condition 
$p_{X}=0$, i.e., $cY=1/2$.
{}From Eq.~(\ref{xcY}) we find $1/3<cY_1<1/2$ and 
$cY_2>1/2$. In the limit $\lambda \to 0$ both fixed points 
approach the phantom divide at $cY=1/2$ (i.e., $w_\phi=-1$).

\subsection{Scaling solutions}

When $\Omega_{\phi}$ is given by (\ref{scaome}), 
Eq.~(\ref{xexpress}) leads to 
\begin{eqnarray}
\label{xscale}    
x=\frac{\sqrt{6}(1+w_{m})}{2(Q+\lambda)}\,.
\end{eqnarray}
This is equivalent to Eq.~(\ref{scarelation}).
{}From Eqs.~(\ref{ws}) and (\ref{scaome}) we 
obtain the effective equation of state given in (\ref{weff2}).
Equations (\ref{basic2}) and (\ref{xscale}) give
\begin{eqnarray}
\label{Omescale}
\Omega_\phi=\frac{Q(Q+\lambda)+3(1+w_{m})p_X}
{(Q+\lambda)^2}\,.
\end{eqnarray}

{}From Eqs.~(\ref{xscale}) and (\ref{weff2}) we find that 
$x$ and $w_{\rm eff}$ are independent of the form of 
$g(Y)$. Meanwhile $Y$ and $y$ depend upon the models
of dark energy.
{}From Eqs.~(\ref{basic1}) and (\ref{xscale}) we find 
\begin{eqnarray}
\label{gderive}    
(1-w_m)g(Y)-2w_{m}Yg'(Y)=
-\frac{2Q(Q+\lambda)}{3(1+w_{m})}\,.
\end{eqnarray}
Once the function $g(Y)$ is specified, one gets $Y$ 
and $y=x/\sqrt{Y}$ by using this equation.
Then we obtain $\Omega_{\phi}$ and $w_{\phi}$
from Eqs.~(\ref{Omescale}) and (\ref{Omew}).

For example, in the case of an ordinary scalar field with 
$g(Y)=1-c/Y$, we obtain 
\begin{eqnarray}
Y&=&\frac{3c(1+w_m)^2}{2Q(Q+\lambda)+3(1-w_{m}^2)}\,, \\
y&=&\left[\frac{2Q(Q+\lambda)+3(1-w_{m}^2)}
{2c(Q+\lambda)^2}
\right]^{1/2}\,,
\end{eqnarray}
which agrees with the result obtained in Ref.~\cite{BNST}.

\section{Stabilities of fixed points}
\label{stability}

In this section we shall study the stability of critical points
obtained in the previous section. Remarkably this can be 
carried out without specifying the functional form of $g(Y)$.
Let us consider small perturbations $\delta x$ and $\delta y$
about the critical point $(x_c, y_c)$, i.e., 
\begin{eqnarray}
x=x_c+\delta x\,, \quad
y=y_c+\delta y\,,\quad
Y=Y_c+\delta Y\,.
\end{eqnarray}
At linear level $\delta Y$ is related to $\delta x$ and 
$\delta y$ via
\begin{eqnarray}
\delta Y=2 \left( \frac{x_{c}}{y_{c}^2}\delta x-
\frac{x_{c}^2}{y_{c}^3}\delta y \right)\,.
\end{eqnarray}

The function $g(Y)$ is expanded as
\begin{eqnarray}
\label{expand}    
g(Y)=g_{c}+g_{c}'(Y-Y_c)+\frac{g_{c}''}{2}
(Y-Y_{c})^2+\cdots\,,
\end{eqnarray}
where $g_{c} \equiv g(Y_{c})$.
When we consider linear perturbations of $g(Y)$,
it is sufficient to neglect the terms higher than 
the first order.
However when we evaluate linear perturbations of 
the $g+Yg'$ term in Eq.~(\ref{xeq}), 
we need to take into account the 
second-order term in Eq.~(\ref{expand}), i.e., 
\begin{eqnarray}
\delta(g+Yg')=\left(2g_c'+Y_cg_c''\right) \delta Y\,.
\end{eqnarray}
Similarly the perturbation of the fractional energy density,
$\Omega_\phi=x^2(g+2Yg')$, is given by 
\begin{eqnarray}
\delta \Omega_\phi=2\frac{x_c}{A_{c}} \delta x
-2\frac{x_{c}^2}{y_{c}}
(3Y_cg_c'+2Y_c^2g_{c}'')\delta y\,,
\end{eqnarray}
where $A_{c}=\left[g_c+5Y_cg_c'+2Y_c^2g_c''\right]^{-1}$.

Let us study linear perturbations about a scalar-field
dominant solution characterized by Eq.~(\ref{fixedi}) and 
a scaling solution characterized by Eq.~(\ref{scaome}).
We recall that these critical points satisfy the equations 
(\ref{basic1}) and (\ref{basic2}).
Then from Eqs.~(\ref{xeq}) and (\ref{yeq})
we get the following perturbation equations:
\begin{eqnarray}
\frac{\rd }{\rd N}
\left(
\begin{array}{c}
\delta x \\
\delta y
\end{array}
\right) = {\cal M} \left(
\begin{array}{c}
\delta x \\
\delta y
\end{array}
\right) \,,
\label{uvdif}
\end{eqnarray}
where the elements of the matrix ${\cal M}$ are
\begin{eqnarray}
\label{a11}
a_{11} &=& -3+\frac{\sqrt{6}}{2}(2Q+\lambda)
x_{c}+3x_c^2(g_c+Y_cg_c') \nonumber \\
& & -3w_m A_c^{-1}x_{c}^2\,,\\
\label{a12}
a_{12} &=& y_{c} \Biggl[ -3g_{c}'x_{c}Y_{c}^2
+\frac{3x_{c}}{y_{c}^2}-\sqrt{6}(Q+\lambda)Y_c \nonumber \\
& &\,\,\,\,\,\,\,\,\,+3w_m x_{c} Y_c 
(A_{c}^{-1}-g_{c}-2Y_{c}g_{c}') 
\nonumber \\
& &
\,\,\,\,\,\,\,\,\,
+\sqrt{6}A_{c} \frac{(Q+\lambda)\Omega_{\phi}+Q}
{2y_{c}^2} \Biggr]\,,\\
\label{a21}
a_{21}&=&\frac{y_{c}}{2} \Biggl[-\sqrt{6}\lambda
+6x_c \left\{(1-w_{m})g_c+(1-5w_m)Y_cg_c' \right\}
\nonumber \\
& &\,\,\,\,\,\,\,\,\,\,-12w_{m}x_cY_c^2g_{c}''  \Biggr]
\,, \\
\label{a22}
a_{22} &=& 3g_{c}'x_c^2Y_c (3w_{m}-1)
+6w_m x_c^2Y_c^2 g_c''\,.
\end{eqnarray}
We checked that these agree what was obtained in 
Ref.~\cite{BNST} for three dark energy models.

The eigenvalues of the matrix ${\cal M}$ are 
\begin{eqnarray}
 \label{deter}   
\mu_{\pm}=\frac{a_{11}+a_{22}}{2}
\left[1 \pm \sqrt{1-\frac{4(a_{11}a_{22}-a_{12}a_{21})}
{(a_{11}+a_{22})^2}} \right]\,.
\end{eqnarray}
The fixed point is stable against perturbations when both 
$\mu_+$ and $\mu_-$ are negative or when $\mu_{\pm}$
have negative real parts. Meanwhile it is unstable
when either $\mu_+$ or $\mu_-$ is positive.

In what follows we study the case of a non-relativistic
barotropic fluid ($w_{m}=0$).
We shall consider the stability of fixed points for 
(i) the scalar-field dominant solution and 
(ii) the scaling solution, separately.

\subsection{Scalar-field dominant solutions}

When $\Omega_{\phi}=1$ we have $(g_c+Y_cg_c')x_c=\lambda/\sqrt{6}$
from Eq.~(\ref{basic2}). Then we find $a_{21}=0$ from 
Eq.~(\ref{a21}). This means that the eigenvalues of 
the matrix ${\cal M}$ are 
\begin{eqnarray}
& &\mu_{+} =a_{11}=-3+\sqrt{6}(Q+\lambda)x_c\,, \\
& &\mu_{-} =a_{22}=-3+
\frac{\sqrt{6}}{2}\lambda x_c\,.
\end{eqnarray}
Making use of Eq.~(\ref{eqw}), the second eigenvalue is expressed 
as $\mu_{-}=-(3/2)(1-w_{\phi})$. 
Hence one has $\mu_{-}<0$ for $w_{\phi}<1$.
The first eigenvalue $\mu_+$ is negative for 
\begin{eqnarray}
\label{scacon}    
x_c<\frac{\sqrt{6}}{2(Q+\lambda)}\,.
\end{eqnarray}
Since we are considering the case of positive values of $Q$
and $\lambda$, $\mu_{-}$ becomes automatically negative
when the condition (\ref{scacon}) is satisfied.
Hence the scalar-field dominant fixed point ($\Omega_{\phi}=1$)
is stable for $x_c<\sqrt{6}/2(Q+\lambda)$, 
whereas it is unstable for 
$x_c>\sqrt{6}/2(Q+\lambda)$.

In the case of a non-phantom scalar field characterized by $p_{X}>0$, 
the stability condition (\ref{scacon}) can be written as
\begin{eqnarray}
\label{sta}    
p_X>\frac{\lambda (Q+\lambda)}{3}\,,
\end{eqnarray}
where we used $p_X x_c=\lambda/\sqrt{6}$.
Meanwhile for a phantom field this stability condition 
corresponds to $p_X<\lambda (Q+\lambda)/3$.
Since $p_{X}<0$ this is automatically satisfied for positive 
values of $Q$ and $\lambda$ and also for $Q=0$ with 
any $\lambda$.
Hence the scalar-field dominant solution is stable 
for phantom fields. This was found in Ref.~\cite{BNST}
for several models of dark energy, but we showed that this 
property persists for any form of $g(Y)$.

\subsection{Scaling solutions}

The scaling solution satisfies Eq.~(\ref{xscale}).
When $w_{m}=0$ we have the relation 
$g_cx_c^2=-Q/(Q+\lambda)$ and also 
$\Omega_\phi=-Q/(Q+\lambda)+2Y_c g_c' x_c^2$.
Hence we find that the components of the 
matrix ${\cal M}$ are 
\begin{eqnarray}
& &
a_{11}=-\frac32 (1-\Omega_{\phi})\,,
\quad
a_{12}=\frac{x_c^2-A_c}{x_{c}y_{c}}a_{22}\,, 
\nonumber \\
& &
a_{21}=\frac{y_{c}}{x_{c}}a_{11}\,,
\quad
a_{22}=-3Y_c g_c' x_c^2\,.
\end{eqnarray}

{}From Eq.~(\ref{deter}) we get the eigenvalues of the 
matrix ${\cal M}$, as
\begin{eqnarray}
\mu_{\pm}=\xi_1 \left[1 \pm \sqrt{1-\xi_2} \right]\,,
\end{eqnarray}
where 
\begin{eqnarray}
& &\xi_1=-\frac{3(2Q+\lambda)}{4(Q+\lambda)}\,, \\
& & \xi_2=\frac{8(1-\Omega_\phi)(Q+\lambda)^3
[\Omega_{\phi} (Q+\lambda)+Q]}{3(2Q+\lambda)^2}A_c\,.
\end{eqnarray}
This shows that $\xi_1<0$ for positive 
$Q$ and $\lambda$. When $Q=0$, $\xi_1$ is 
automatically negative ($\xi_1=-3/4$).
$\mu_+$ is negative or has a negative real part.
Meanwhile the sign of $\mu_-$ depends upon 
that of $\xi_2$.
When $\xi_2>0$ the scaling solution is stable, whereas 
it is unstable for $\xi_2<0$.

In order to get a viable scaling solution we require the 
condition $\Omega_\phi<1$.
Hence the stability of the scaling solution is dependent 
on the sign of $A_c$, which can be expressed as
\begin{eqnarray}
 \label{Ac}
A_c=(p_{X_c}+2X_{c}p_{X_cX_c})^{-1}\,.
\end{eqnarray}
This quantity is related to the speed of sound:
\begin{eqnarray}
\label{sound}
c_s^2 \equiv \frac{p_X}{\rho_X}=
\frac{p_X}{p_X+2Xp_{XX}}\,,
\end{eqnarray}
which appears as a coefficient of the  
$k^2/a^2$ term in perturbation equations 
\cite{Garriga,ATS}
($k$ is a comoving wavenumber).

While the classical fluctuations may be
regarded to be stable when $c_s^2>0$, it was shown in 
Ref.~\cite{PT} that the stability of 
{\it quantum} fluctuations requires the conditions
$p_{X}>0$ and $p_X+2Xp_{XX}>0$.
Hence if we impose the quantum stability, $A_c$
is positive. Then the scaling solution with $\Omega_{\phi}<1$
is a stable attractor.
In the case of an ordinary scalar field with $p=X-V(\phi)$
one has $A_c=1$, which shows that the scaling solution
is stable. Meanwhile for an ordinary phantom field 
with $p=-X-V(\phi)$, it is unstable since $A_c=-1$.

{}From Eq.~(\ref{Omescale}) with $w_{m}=0$ 
we find that the condition, $\Omega_{\phi}<1$, 
corresponds to 
\begin{eqnarray}
\label{pxcon}    
p_X<\frac{\lambda (Q+\lambda)}{3}\,.
\end{eqnarray}
When this is satisfied, the stability condition (\ref{sta})
for the scalar-field dominant fixed point is violated.
Hence when we have a stable scaling solution with 
$\Omega_{\phi}<1$ and $A_c>0$, the scalar-field 
dominant fixed point with $p_{X}>0$ is always unstable.
Hence the system chooses 
either a scaling solution
with $\Omega_{\phi}<1$ or a scalar-field dominant
solution with $\Omega_{\phi}=1$ and $p_{X}>0$
as an attractor. In Ref.~\cite{BNST} this property was 
found by specifying the functional form of $g(Y)$, 
but we have shown that this holds 
for all scalar-field models which possess scaling solutions.

We note that the scaling solution can be an attractor
even for a phantom ($p_{X}<0$) provided that the
condition $p_X+2Xp_{XX}>0$ is satisfied.
In the case of the dilatonic ghost condensate model 
($p=-X+ce^{\lambda \phi}X^2$), for example, 
this is realized for $1/3<cY<1/2$. 
The scalar-field dominant solution 
$cY_{1}=1/2+\lambda^2 f_-(\lambda)/16$
in Eq.~(\ref{xcY}) also exists in the 
phantom region characterized by $1/3<cY_{1}<1/2$.
In the subsection A we showed that 
this case is also a stable attractor.
We recall that $x_1$ in Eq.~(\ref{xcY}) is negative, 
whereas $x$ in Eq.~(\ref{xscale}) is positive.
We have numerically found that the solutions approach
the scaling solution for the initial condition $x_{i}>0$
whereas they approach the phantom scalar-field dominant
point when $x_{i}<0$. 
Note that another scalar-field dominant solution 
$cY_{2}=1/2+\lambda^2 f_+(\lambda)/16$ ($>1/2$)
in Eq.~(\ref{xcY}) is unstable when the scaling 
solution is stable.
We caution that the phantom field ($p_X<0$)
with $p_X+2Xp_{XX}>0$ corresponds to a negative
sound of speed, thus suffering from the instability of 
perturbations even at the classical level \cite{PT}.
See Refs.~\cite{dilaton,Vikman} for related works.

\section{Multi-field dark energy}

We can generalize the analysis to the case in which 
many fields are present with a same Lagrangian 
density for the each field. 
In the case of an ordinary scalar field with an 
exponential potential several fields can cooperate to 
drive an accelerated expansion even if none is able to do 
individually \cite{LMS,Coley,KLT}.
We will show that this generally happens for the 
Lagrangian density which possesses 
cosmological scaling solutions. 
We shall study the case without a coupling 
between dark energy and dark matter ($Q=0$)
throughout this section. This is because an accelerated
expansion is realized even for $Q=0$ in the presence 
of multiple scalar fields.

Let us consider $n$ scalar fields 
($\phi_{1}, \phi_{2}, \cdots, \phi_n$) with 
the Lagrangian density:
\begin{eqnarray}
p=\sum_{i=1}^n X_i g(X_i e^{\lambda_i \phi_i})\,,
\end{eqnarray}
where $X_{i} =-g^{\mu\nu} \partial_\mu \phi_i 
\partial_\nu \phi_i/2$ and
$g$ is an arbitrary function.
In the presence of a barotropic fluid with an 
equation of state $w_m=p_m/\rho_m$, 
the constraint equation is given by 
$\Omega_m+\Omega_\phi=1$ 
with a fractional density
\begin{eqnarray}
\Omega_{\phi}=\sum_{i=1}^n \Omega_{\phi_i}
=\sum_{i=1}^n x_i^2 \left[g(Y_i)+2Y_ig'(Y_i)
\right]\,.
\end{eqnarray}
We also obtain the equation for the Hubble rate:
\begin{eqnarray}
\frac{1}{H} \frac{{\rm d}H}{{\rm d}N}
&=&-\frac32 \Biggl[1+\sum_{i=1}^n g(Y_i)x_i^2-w_{m}
(\Omega_\phi-1) \Biggr]\,.
\end{eqnarray}

Each scalar field, $\phi_{i}$, satisfies the equation 
of motion (\ref{ddphi}) with $Q=0$.
Defining dimensionless quantities
$x_{i} \equiv \dot{\phi}_{i}/\sqrt{6}H$ and 
$y_{i} \equiv e^{-\lambda_{i}\phi_i/2}$, we
get the following differential equations:
\begin{eqnarray}
\frac{{\rm d}x_i}{{\rm d}N}&=&\frac{3x_{i}}{2}
\left[1+\sum_{i=1}^n g(Y_i)x_i^2-w_{m}
(\Omega_\phi-1)-\frac{\sqrt{6}}{3}
\lambda_{i}x_{i}\right] \nonumber \\
&&+\frac{\sqrt{6}A}{2} \left[\lambda_{i} 
\Omega_{\phi_i}-\sqrt{6}\{g(Y_i)+Y_ig'(Y_i)\}
x_{i} \right]\,, \\
\frac{{\rm d}y_i}{{\rm d}N}&=&\frac{3y_i}{2}
\left[1+\sum_{i=1}^n g(Y_i)x_i^2-w_{m}
(\Omega_\phi-1)-\frac{\sqrt{6}}{3}\lambda_{i}
x_{i} \right]\,. \nonumber \\
\end{eqnarray}
Then the fixed points we are interested 
in ($y_i \neq 0$) satisfy 
\begin{eqnarray}
\label{lam1}    
\lambda_i x_i &=&
\frac{\sqrt{6}[g(Y_i)+Y_ig'(Y_i)]}
{g(Y_i)+2Y_ig'(Y_i)} \\
&=& 
\frac{\sqrt{6}}{2}
\left[1+\sum_{i=1}^n g(Y_i)x_i^2-w_{m}
(\Omega_\phi-1) \right]\,.
\label{lam2d}   
\end{eqnarray}

Since the equation of state of the each scalar field is 
given by $w_{\phi_i}=
g(Y_i)/[g(Y_i)+2Y_ig'(Y_{i})]$, we find 
\begin{eqnarray}
\frac{g(Y_i)+Y_ig'(Y_{i})}{g(Y_i)+2Y_ig'(Y_{i})}
=\frac12 (1+w_{\phi_i})\,.
\end{eqnarray}
Then from Eq.~(\ref{lam1}) we get
\begin{eqnarray}
\label{multieff}    
w_{\phi_i}=-1+\frac{\sqrt{6}}{3}
\lambda_i x_i\,.
\end{eqnarray}
Equation (\ref{lam2d}) indicates that 
the quantities $\lambda_i x_i$
are independent of $i=1, 2, \cdots, n$. 
Hence we set 
\begin{eqnarray}
\label{lamx}
\lambda_1 x_1=\cdots=\lambda_i x_i=
\cdots =\lambda_n x_n \equiv \lambda x\,.
\end{eqnarray}

Then from Eq.~(\ref{multieff}), $w_{\phi_{i}}$
are same for all scalar fields, i.e., 
\begin{eqnarray}
\label{wphii}    
w_{\phi_1}=\cdots=w_{\phi_{i}}=
\cdots =w_{\phi_{n}} \equiv w_{\phi}\,.
\end{eqnarray}
{}From Eq.~(\ref{lam1}) the quantities $Y_i$ are determined 
by $\lambda_i x_i$, which means that $Y_i$ are also  
same independent of $i$:
\begin{eqnarray}
\label{Yi}
Y_1=\cdots=Y_i=
\cdots =Y_n \equiv Y\,.
\end{eqnarray}

Taking note the relation 
\begin{eqnarray}
 \label{relation}   
\sum_{i=1}^n g(Y_i) x_i^2
=\sum_{i=1}^n w_{\phi_i}\Omega_{\phi_i}
=w_{\phi} \Omega_\phi\,,
\end{eqnarray}
we find that Eq.~(\ref{lam2d}) yields
\begin{eqnarray}
(\Omega_{\phi}-1)(w_\phi- w_m)=0\,.
\end{eqnarray}
Hence we have two fixed points: (i) scalar-field 
dominant solution with $\Omega_{\phi}=1$
and (ii) scaling solution with $w_\phi=w_m$.
In what follows we shall discuss 
these cases separately.

\subsection{Scalar-field dominant solutions ($\Omega_\phi=1$)
and assisted dark energy}

{}From Eqs.~(\ref{lamx}) and (\ref{Yi}), 
Eq.~(\ref{lam1}) is written as
\begin{eqnarray}
\label{single1}    
\lambda x =
\frac{\sqrt{6}[g(Y)+Yg'(Y)]}
{g(Y)+2Yg'(Y)}\,.
\end{eqnarray}
Using Eq.\,(\ref{lamx}) with $\Omega_{\phi}=1$,
we find that Eq.~(\ref{lam2d}) is given by 
\begin{eqnarray}
\label{lambdax}    
\lambda x=\frac{\sqrt{6}}{2}
\left[ 1+g(Y) x^2\lambda^2 
\sum_{i=1}^n \frac{1}{\lambda_i^2}
\right]\,.
\end{eqnarray}
Hence if we choose 
\begin{eqnarray}
\label{efflambda}    
\frac{1}{\lambda^2}=
\sum_{i=1}^n \frac{1}{\lambda_i^2}\,,
\end{eqnarray}
Equation (\ref{lambdax}) yields
\begin{eqnarray}
\label{single2}  
\lambda x=\frac{\sqrt{6}}{2}
\left[ 1+g(Y) x^2\right]\,.
\end{eqnarray}

Equations (\ref{single1}) and (\ref{single2}) show that 
the system effectively reduces to that of the 
single field with $\lambda$ given by Eq.~(\ref{efflambda}).
{}From Eq.~(\ref{single1}) and $\Omega_\phi=1$
the quantity $p_{X}=g(Y)+Yg'(Y)$ 
satisfies the relation $\sqrt{6}xp_{X}=\lambda$.
Hence from Eq.\,(\ref{multieff}) with Eqs.\,(\ref{lamx})
and (\ref{wphii}), we get the equation of state:
\begin{eqnarray}
\label{weff}    
w_{\phi}=-1+\frac{\lambda^2}{3p_X}\,,
\end{eqnarray}
which is the same expression as Eq.~(\ref{wsingle}).

Let us consider the case of a non-phantom scalar 
field ($p_{X}>0$). Equation (\ref{efflambda})
shows that the presence of multiple scalar fields leads to 
the decrease of $\lambda^2$ relative to the single-field
case. The quantity $Y$ is known by Eq.~(\ref{single1})
in terms of $\lambda x$ once $g(Y)$ is specified.
Then $p_{X}(Y)=g(Y)+Yg'(Y)$ is determined by 
the quantity $\lambda x$ which is unaffected 
by adding scalar fields.
This means that the effect of 
multiple scalar fields to the equation of state 
only appears for $\lambda^2$ in Eq.~(\ref{weff}).
This effect works to reduce the equation of state toward 
$w_{\phi}=-1$ as $\lambda$ decreases.
Hence even if inflation does not occur in the single-field case
because of large $\lambda$, the presence of many scalar fields
can lead to an accelerated expansion by reducing $\lambda$.
Thus we have shown that assisted acceleration occurs
for all dark energy models which possess scaling solutions.

In the case of a phantom field ($p_{X}<0$)
the presence of many scalar fields leads to 
the {\it increase} of $w_{\phi}$ toward $-1$, which is 
different from the assisted acceleration.
In any case the equation of state approaches that of 
cosmological constant ($w_{\phi}=-1$) as we add 
more scalar fields.

Taking note the relation $g(Y_i)x_{i}^2=w_{\phi_{i}}
\Omega_{\phi_{i}}$ and $g(Y)x^2=w_{\phi}$ together 
with (\ref{wphii}) and (\ref{Yi}), the fractional energy 
density of each scalar field is 
\begin{eqnarray}
\Omega_{\phi_{i}}=\frac{x_i^2}{x^2}
=\frac{\lambda^2}{\lambda_i^2}\,.
\end{eqnarray}
In the two-field case, for example, one has 
$\Omega_{\phi_1}=\lambda_2^2/(\lambda_1^2+\lambda_2^2)$
and $\Omega_{\phi_2}=\lambda_1^2/(\lambda_1^2+\lambda_2^2)$.
The energy density is distributed so that the condition 
(\ref{lamx}) is satisfied.

\subsection{Scaling solutions}

The scaling solution satisfies $w_{\phi_{i}}=w_{\phi}=w_{m}$
($i=1, 2, \cdots, n$). 
Hence we do not have an accelerated expansion for a non-relativistic 
dark matter ($w_{m}=0$).
{}From Eq.~(\ref{lam2d}) and (\ref{relation}) we obtain
\begin{eqnarray}
\label{xi}
x_{i}=\frac{\sqrt{6}(1+w_{m})}{2\lambda_i}\,.
\end{eqnarray}
This is similar to the single field case, see
Eq.~(\ref{xscale}) with $Q=0$.
{}From Eq.~(\ref{lam1}) one gets 
$\Omega_{\phi_i}=\sqrt{6}x_{i}p_X/\lambda_{i}$.
Then by using Eq.~(\ref{xi}) we find 
$\Omega_{\phi_i}=3(1+w_m)p_X/\lambda_i^2$.
This gives the total energy fraction of the scalar field:
\begin{eqnarray}
\Omega_{\phi}=\sum_{i=1}^n \Omega_{\phi_i}
=\frac{3(1+w_{m})p_X}{\lambda^2}\,,
\end{eqnarray}
where $\lambda$ is defined in Eq.~(\ref{efflambda}).
In the case of an ordinary scalar field with $p_{X}=1$,
this result agrees with what was obtained in Ref.~\cite{KLT}.

In order for the scaling solution to be physically meaningful,
we require the condition $\Omega_\phi<1$, i.e., 
\begin{eqnarray}
p_{X}<\frac{\lambda^2}{3(1+w_{m})}\,.
\end{eqnarray}
In the case of a non-relativistic dark matter 
this is equivalent to the condition (\ref{pxcon}) with $Q=0$.
Since the multi-field system is now reduced to an effectively 
single-field system, the stability analysis of critical points
discussed in Sec.\,\ref{stability} persists 
in the presence case as well. 
Thus the scaling solution  is stable 
if a scalar-field dominant solution with $p_{X}>0$
is unstable, and vice versa. 
This is actually what was found in Ref.~\cite{KLT} 
for an ordinary field with an exponential potential.

As discussed in Ref.~\cite{KLT} we can consider an interesting 
situation in which the solution is in the scaling regime 
characterized by $w_{\phi_{i}}=w_{m}$ during radiation and 
matter dominant eras and then enters the regime of an accelerated 
expansion as more and more fields join the assisted scalar-field 
dominant attractor. Although this requires a fine-tuning, it is interesting that 
the presence of many scalar fields leads to the possibility 
of an accelerated expansion.

\section{Conclusions}

In this paper we have provided a general  method to derive critical 
points relevant to dark energy for the models which possess
scaling solutions. The existence of scaling solutions restricts 
the Lagrangian density to be in the form $p=Xg(Xe^{\lambda \phi})$
where $g$ is an arbitrary function.
This includes a wide variety of dark energy models such as 
quintessence, phantoms, tachyons and dilatonic ghost condensates. 
Since $\lambda$ is constant, we obtain autonomous equations
(\ref{xeq}) and (\ref{yeq}) for two variables $x$ and $y$
defined in Eq.\,(\ref{xydef}).
For the models in which scaling solutions do not exist, one has another
differential equation for $\lambda$ since $\lambda$ is a dynamically 
varying quantity. Even in this case we can apply the results 
obtained in this paper by considering ``instantaneous'' 
critical points $(x(N), y(N))$ \cite{Maco}.

The fixed points relevant to dark energy is (i) the scalar-field 
dominant solution ($\Omega_{\phi}=1$) and 
(ii) the scaling solution with 
$\rho_{\phi}/\rho_{m}={\rm constant} \neq 0$. 
In the former case the scalar-field equation of state is generally 
given by Eq.~(\ref{eqw}). The fixed points can be derived by 
using Eq.~(\ref{gre1}) when we specify the function $g(Y)$.
Depending upon the models of dark energy the fixed points 
exist in the region given by $p_{X}>0$ or $p_{X}<0$.
In some of the models like dilatonic ghost condensate
there are two fixed points both in the regions $p_X>0$
and $p_{X}<0$.
In the case of scaling solutions the variable $x$ and the effective 
equation of state $w_{\rm eff}$ are uniquely determined as 
Eqs.~(\ref{xscale}) and (\ref{weff2}) independently of
the models. The variables $y$ and $Y=x^2/y^2$ are known 
by Eq.~(\ref{gderive}) once we specify the form of $g(Y)$.

The stability of fixed points was discussed in Sec.~\ref{stability}
in the presence of a non-relativistic barotropic fluid
without specifying any form of $g(Y)$.
The stability condition for the scalar-field 
dominant solution is given by Eq.~(\ref{sta}) for 
a non-phantom field ($p_{X}>0$).
When this condition is satisfied, the scaling solution 
exists in the region which is not physically meaningful 
($\Omega_{\phi}>1$).
When the quantity $A_c$ defined in Eq.~(\ref{Ac})
is positive, the scaling solution is a stable attractor 
for $\Omega_{\phi}<1$. The positivity of $A_{c}$
is ensured if we impose the stability of quantum fluctuations.
Hence when the stable scaling solution exists in the region 
$\Omega_{\phi}<1$, the scalar-field dominant solution 
with $p_X>0$ is unstable.

The above discussion shows that the solution chooses either the
scalar-field dominant solution with $p_X>0$ or the scaling 
solution with $\Omega_{\phi}<1$.
When there exists a phantom scalar-field dominant 
fixed point ($p_X<0$), we found that this solution is 
classically stable. If both the scaling solution and this
phantom scalar-field dominant solution are stable critical 
points, the solutions choose either of them depending 
upon initial conditions. 

Finally we generalized our analysis to the case in which more than 
one field is present with the same form of the Lagrangian density.
We showed that the system effectively reduces to that of the 
single-field with $\lambda$ modified as Eq.~(\ref{efflambda})
for the scalar-field dominant solution. Since the presence of 
many scalar fields leads to the decrease of $\lambda$ toward 0, 
the equation of state approaches that of cosmological constant
as we add more fields.
Thus we have shown that for a non-phantom fluid the assisted
acceleration occurs for all scalar-field dark energy 
models which have scaling solutions.

It is really remarkable that the derivation of critical points,
their stability and assisted acceleration can be analyzed 
in a unified way without specifying any form of $g(Y)$.
We hope that the results obtained in this paper will be 
useful to constrain scalar-field dark energy models from 
future high-precision observations.
It is certainly of interest to place strong observational constraints
not only from SN Ia but from CMB and LSS 
by using the equations of matter perturbations
derived in Refs.~\cite{ATS}.

\section*{ACKNOWLEDGMENTS}
This work is supported by JSPS
(Grant No.\,30318802).

\end{document}